\documentclass[12pt]{article}

\usepackage{graphicx}

\begin{document}

\begin{titlepage}

%\rightline{hep-th/0606249}

%\vskip 2cm

\centerline{\large \bf {Cosmological constraints on a dark energy}}
\centerline{\large \bf {model with a non-linear scalar field}}

\vskip 1cm

\centerline{G. Panotopoulos}

\vskip 1cm

\centerline{Department of Physics, University of Crete,}

\vskip 0.2 cm

\centerline{Heraklion, Crete, GREECE}

\vskip 0.2 cm

\centerline{email:{\it panotop@physics.uoc.gr}}

\begin{abstract}
In the present work we study a dark energy model in which a non-linear scalar field (tachyon) with a Born-Infeld type of
action is responsible for the observed cosmic acceleration. The potential of the tachyon is well-motivated since it
comes from open string theory and the model is subjected to various cosmological constraints with data coming from
supernovae as well as from microwave background radiation. Our analysis shows that in the particular model under study the tachyon can be
an excellent candidate for dark energy in the universe, as the model agrees with a series of observational
data and for a wide range of the parameters of the model.
\end{abstract}

\end{titlepage}

\section{Introduction}

Perhaps the most dramatic discovery in modern cosmology is the observational fact that the present universe undergoes an
accelerating phase. During the last ten years or so a remarkable progress in technology has allowed us to witness extraordinary
precision measurements in cosmology. A plethora of observational data are now available, which show that we live in a
flat universe that expands with an accelerating rate and that the dominant component in the energy budget of the universe
is an unusual material, the nature of which still remains unknown. Identifying the origin and nature of dark energy is
one of the great challenges in modern theoretical cosmology. The simplest candidate for dark energy is the cosmological
constant, which corresponds to a perfect fluid with parameter $w=p/\rho=-1$. However, over the years many other
theoretical models have been proposed and studied. One class of such models is based on some modification of Einstein's
gravity~\cite{gde1,gde2} and one is talking about the so called geometrical dark energy models. Another class contains the dynamical dark
energy models, in which a new dynamical field (almost certainly a scalar field) is coupled to gravity. In the second
class one would find models called quintessence~\cite{quintessence}, phantom~\cite{phantom}, quintom~\cite{quintom}, k-essence~\cite{kessence}, tachyonic~\cite{tachyonic} etc. A recent review on dark energy dynamics one can find in~\cite{copeland}.

Since Sen's original publications~\cite{sen}, much work has been done on tachyonic cosmology, concerning either
inflation~\cite{tachinfl} or dark
energy~\cite{tachyonic}. In the present work our aim is to test a particular tachyonic dark energy model against
observational
cosmological data that are available from supernovae and microwave background radiation. A very recent work~\cite{liddle} is closely related to ours and hopefully the present analysis will be a complementary one, at least for the particular model that we shall study here. In~\cite{liddle} the authors studied a wide range of different tachyonic models. In one of these models the
potential for the tachyon emerges from a class of open string theories (both bosonic and superstrings) with explicit computation.
That way string theory can be linked to cosmology and provide a natural candidate for playing the role of dark energy in the universe. This is exactly the potential that we shall be using in what follows.

The present work is organized as follows: There are four sections of which this introduction is the first. The theoretical framework is presented in section 2, while our analysis and results are presented in section 3. Finally we conclude in the fourth section.

\section{The theoretical framework}

We work in the framework of a four-dimensional FRW cosmology, considering the case of a flat ($k=0$) universe. The
Friedmann equations have the well-known form
\begin{eqnarray}
H^2 & = & \frac{8 \pi G}{3} \: (\rho_m+\rho_X) \\
\frac{\ddot{a}}{a} & = & -\frac{4 \pi G}{3} \: (\rho_m+\rho_X+3 p_X)
\end{eqnarray}
where $G$ is Newton's constant, $a$ is the scale factor, $H=\dot{a}/a$ is the Hubble parameter, $\rho_X, p_X$ are the
energy density and pressure for dark energy respectively and $\rho_m$ is the energy density for the pressureless (dust)
non-relativistic matter, $p_m = 0$. We define the critical density $\rho=(3H^2)/(8 \pi G)$ and the normalized densities
for dark energy and matter $\Omega_m=\frac{\rho_m}{\rho_{cr}}$ and $\Omega_X=\frac{\rho_X}{\rho_{cr}}$. From the first
Friedmann equation one can easily see that $\Omega_m+\Omega_X=1$. In cosmology we often use instead of time or
temperature the red-shift, $z=-1+a_o/a$, where $a_o$ is the present value of the scale factor. In general we use the
sub-index "$o$" to indicate present values of various quantities. In an isotropic and homogeneous universe all quantities
are functions of time only, and so is for the red-shift and the Hubble parameter. Eliminating time we can obtain a
formula relating the Hubble parameter to the red-shift~\cite{copeland}
\begin{equation} \label{1}
H(z) = H_o \sqrt{\Omega_m (1+z)^3+\Omega_X f(z)}\equiv H_o E(z)
\end{equation}
where $f(z)$ is a function of the red-shift depending on the model with the property
$f(0)=1$, since $\Omega_m+\Omega_X=1$. $\Omega_m, \Omega_X$ are the present values of the normalized densities for matter
and dark energy respectively, but we drop the index "$o$." For example, if the cosmological constant is dark energy
then $f(z)=1$, or if some fluid with constant parameter $w$ is dark energy then $f(z)=(1+z)^{3 (1+w)}$. In the general
case $f(z)$ is a more complicated expression and is given by~\cite{copeland}
\begin{equation}
f(z)=\textrm{exp} \left ( 3 \int_0^z dx \: \frac{1+w_X(x)}{1+x} \right )
\end{equation}
Equation (\ref{1}) is a key formula for our analysis because from that we
can obtain all the useful expressions, as it will be shown in a while. Since we are interested in the cosmic acceleration
we define the deceleration parameter (the name deceleration was given originally because it was believed that the
universe expands in a decelerating rate) $q=-\ddot{a}/(a H^2)$ and using the second Friedmann equation and the definition
for the red-shift we obtain~\cite{leandros1}
\begin{equation}
q(z)=-1+(1+z) \frac{H'(z)}{H(z)}
\end{equation}
where $H'(z)$ is the derivative of $H(z)$ with respect to $z$. Furthermore, for a generic dark energy, its
equation-of-state parameter $w_X=p_X/\rho_X$ is given by~\cite{leandros1}
\begin{equation}
w_X(z)=\frac{-1+\frac{2}{3} (1+z) \frac{H'(z)}{H(z)}}{1+\frac{H_o^2}{H(z)^2} \Omega_m (1+z)^3}
\end{equation}
The final useful expression that can be computed from (\ref{1}) is the luminosity distance (for distances in
cosmology see~\cite{distances}), which plays a central role for the comparison between data from supernovae and a
theoretical model. For a flat universe the luminosity distance $D_L$ as a function of the red-shift is given by
\begin{equation}
D_L(z)=(1+z) \: \int_0^z dx \: \frac{1}{H(x)}
\end{equation}
Now we assume that the role of dark energy in the universe is played by the tachyon, a scalar field which is coupled to gravity non-linearly with an action
\begin{equation}
S_{BI}=-V(\phi) \int d^4x \sqrt{-g} \sqrt{1-g^{\mu \nu} \partial_{\mu} \phi \partial_{\nu} \phi}
\end{equation}
where $\phi$ is the tachyon and $V(\phi)$ is its potential. In our work we shall consider a potential of the form~\cite{potential}
\begin{equation}
V(\phi)=\frac{V_0}{\textrm{cosh}(\phi/\phi_0)}
\end{equation}
with $V_0, \phi_0$ the two positive parameters of the model.
The equation of motion of a homogeneous tachyon $\phi=\phi(t)$ is
\begin{equation}
\frac{\ddot{\phi}}{1-\dot{\phi}^2}+3 H \dot{\phi}+\frac{V'(\phi)}{V(\phi)}=0
\end{equation}
Similarly to the linear Klein-Gordon scalar field, the tachyon also behaves as a perfect fluid with energy density $\rho_\phi$ and pressure $p_\phi$ that are given by the expressions
\begin{eqnarray}
\rho_\phi & = & \frac{V(\phi)}{\sqrt{1-\dot{\phi}^2}} \\
p_\phi & = & -V(\phi) \sqrt{1-\dot{\phi}^2}
\end{eqnarray}
Notice that the pressure that corresponds to the tachyon is negative. The state parameter $w_\phi$ is given accordingly
\begin{equation}
w_\phi = \frac{p_\phi}{\rho_\phi}=\dot{\phi}^2-1
\end{equation}
and it is a function of time or red-shift if we wish. It is easy to see that the parameter $w_\phi$ takes values in the interval between $0$ and $-1$. One limit comes from the fact that $\dot{\phi}^2 \geq 0$, while the other limit comes from the requirement that $1-\dot{\phi}^2 > 0$. Therefore from this point of view the tachyon is similar to the Klein-Gordon field and it does not cross the phantom devide line. So according to an analysis made in~\cite{leandros2} we expect that our tachyonic dark energy model will not have a better fit to supernovae data compared to the standard LCDM model. In~\cite{leandros2} it has been shown that models which allow $w(z=0) < -1$ and crossing of the phanton devide line $w=-1$ produce better fits to the Gold dataset.

\section{Analysis and results}

Let us start with what we know observationally about dark energy. The first thing about dark energy has to do with its state parameter and the second about how much dark energy there is in the universe. As far as the latter is concerned, dark energy comprises the biggest part of the energy content of the universe, $\Omega_X \sim 0.7$. Also, recent results give a strong support that the present time dark energy state parameter is close to $-1$~\cite{koshelev}
\begin{equation}
w=-0.97_{-0.09}^{+ 0.07}
\end{equation}
or without an a priori assumption that the universe is flat $w=-1.06_{-0.08}^{+ 0.13}$~\cite{koshelev}. In our analysis we shall assume that the universe is flat and that the normalized density for matter is $\Omega_m = 0.3$. The observational data that we shall use to test our model are i) the Gold dataset~\cite{riess}, ii) the CMB shift parameter $S$~\cite{maartens}, iii) a parameter $A$ that is constrained by the observed scale of the first acoustic peak in baryon oscillations~\cite{maartens} and iv) the redshift distance $r(z_{dec})$~\cite{amendola}. In particular, the above quantities are given by the expressions~\cite{maartens, amendola}
\begin{eqnarray}
S & = & \sqrt{\Omega_m} \: H_o \: \frac{D_L(z_r)}{1+z_r}, \quad z_r=1089 \\
A & = & \sqrt{\Omega_m} \: \left ( \frac{H_o^3 D_L(z_1)}{H_1 z_1^2 (1+z_1)^2}  \right )^{1/3}, \quad z_1=0.35 \\
r(z_{dec}) & = & \int_0^{z_{dec}} dz \: \frac{1}{H(z)}, \quad z_{dec}=1100
\end{eqnarray}
and their values are as follows~\cite{maartens, amendola}
\begin{eqnarray}
S & = & 1.716 \pm 0.062 \\
A & = & 0.469 \pm 0.017 \\
r(z_{dec}) & = & (13.7 \pm 0.5)~\textrm{Gpc}
\end{eqnarray}
Our goal now is to integrate the dynamical equations and obtain i) the tachyon state parameter $w_\phi$ as a function of red-shift $z$ and ii) the Hubble parameter $H$ as a function of $z$, from which we then can compute the deceleration parameter $q(z)$ as well as the luminosity distance $D_L(z)$ as we have already mentioned. Recently in~\cite{leandros3} a comparison was made between the SNI Gold dataset and phantom and quintessence models with linear potentials. The mathematical file that the authors of ~\cite{leandros3} used can be found in~\cite{url}. We have used this file modifying it where appropriate depending on our needs. Following~\cite{leandros3} for initial conditions ($t \rightarrow t_i \approx 0$) we use
\begin{eqnarray}
\dot{a}(t_i) & = & \left ( \frac{9}{4} \: \Omega_m \right ) \: t_i^{2/3} \\
\phi(t_i) & = & \phi_i \\
\dot{\phi}(t_i) & = & 0
\end{eqnarray}
The initial value for the tachyon $\phi_i$ is chosen so that the normalized density for dark energy today is $\Omega_X=\Omega_\phi=1-\Omega_m=0.7$. The present time (age of the universe) $t_o$ is defined by $a(t_o)=1=H(t_o)$~\cite{leandros3}.

A plot of $w_\phi$ versus $z$ is shown in figure 1. A comparison between the supernovae data and the theoretical curve is shown in figure 2, while the deceleration parameter in our model compared to that of LCDM model is shown in figure 3. For this case the theoretical values for the CMB shift parameter $S$, the parameter $A$, the redshift distance $r(z_{dec})$ and the present value for the dark energy state parameter $w_\phi$ we have found are
\begin{eqnarray}
S & = & 1.741 \\
A & = & 0.485 \\
r(z_{dec}) & = & 13.6~\textrm{Gpc} \\
{w_\phi}_o & = & -0.96
\end{eqnarray}
in agreement with the observational values. To see if this model produces a better fit compared to LCDM model, one should use the $\chi^2$ method~\cite{leandros3}. For the case shown in figures 1,2,3 the computed $\chi^2$ is $\chi^2=177.4$, while for LCDM $\chi^2=177.1$. This confirms our expectation that tachyon models do not produce better fits to the Gold dataset.  In general, our detailed analysis shows that the behavior of the system is the following. All the theoretical values agree with observations, the theoretical curve for the supernovae apparent magnitude fits rather well the Gold dataset and the deceleration parameter for our model lies very close to the one for LCDM model, as long as $w(z)$ lies in the interval $\sim -0.9$ to $-1$. As it moves to larger values the agreement becomes worse and worse. As an extreme case, we report that when $w \sim -1/3$ ($w(z)$ is shown in figure 4) the deceleration parameter can be seen in figure 5. Notice that after a short period of an accelerated expansion, the universe enters again the deceleration era. This can be understood from the equation for the deceleration parameter
\begin{equation}
q=\frac{\Omega_m}{2}+\frac{\Omega_X}{2} \: (1+3 w_X)
\end{equation}
If today $\Omega_m = 0.3, \Omega_X = 0.7, w_X^{o} \approx -0.33$, then from the equation above one can easily see that $q > 0$. Finally for this extreme case we
report the theoretical values for $S, A, r(z_{dec})$
\begin{eqnarray}
S & = & 1.672 \\
A & = & 0.443 \\
r(z_{dec}) & = & 13.09~\textrm{Gpc} \\
{w_\phi}_o & = & -0.36
\end{eqnarray}
We see that the CMB shift parameter $S$ is still within observational limits, but all the other quantities obviously disagree with the data.

\section{Conclusions}

In the present work we have studied a dark energy model with a tachyon playing the role of dark energy in the
universe. The tachyon is non-linearly coupled to gravity, with a Born-Infeld type of action. Its equation of motion is
different than that of a Klein-Gordon scalar field, which couple linearly to gravity. However, here also the homogeneous
tachyon behaves like a perfect fluid with an energy density and pressure that are given by certain expressions in terms
of the potential and the time derivative of the tachyon. The potential for the tachyon that we have used emerges from
explicit computations in open string theories (bosonic string or superstrings), and therefore is well-motivated. We have
tested the particular model against a series of cosmological constraints coming from observational data both from
supernovae and microwave background radiation. Our detailed analysis shows that in the model under consideration the tachyon can be an
excellent candidate for dark energy, as the model passes successfully all the aforementioned tests for a wide range of the parameters of the model.

\section*{Acknowlegements}

Special thanks to L.~Perivolaropoulos for discussions and for giving us the supernovae Gold dataset and the
mathematica file. We would like also to thank the Physics Department of Technion  for its warm hospitality, where the present work was completed. This work was supported by the EU grant MRTN-CT-2004-512194.

\newpage

\begin{figure}[h!]
\centering
%\hspace{0.1cm}%
\begin{tabular}{cc}
\includegraphics*[width=240pt, height=180pt]{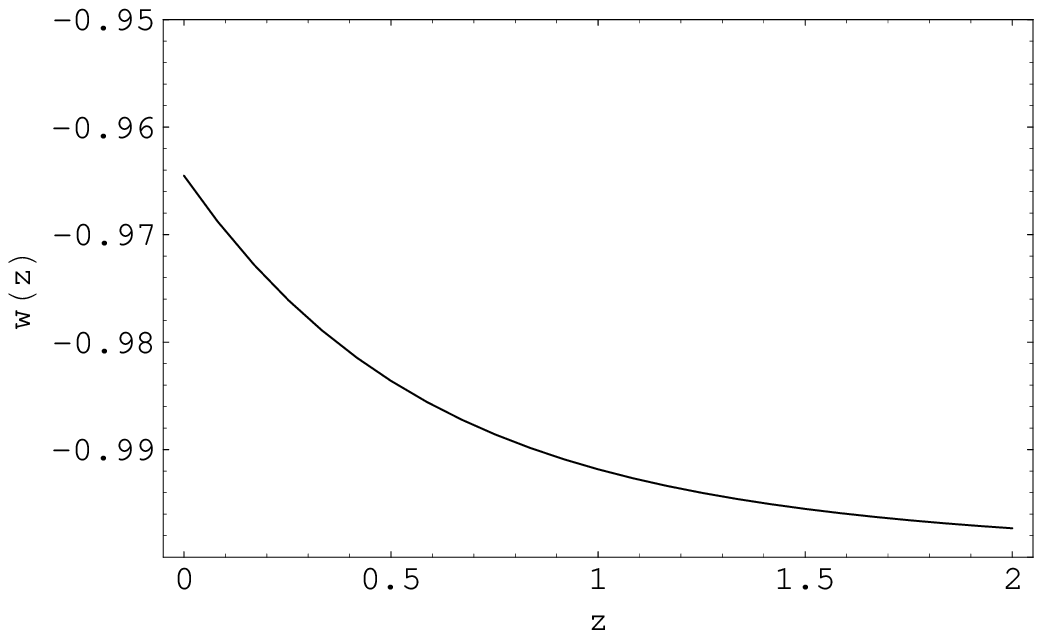}&%
\end{tabular}
 \caption{
State parameter for tachyon $w$ as a function of red-shift $z$.}
\end{figure}

\begin{figure}[h!]
\centering
%\hspace{0.1cm}%
\begin{tabular}{cc}
\includegraphics*[width=240pt, height=180pt]{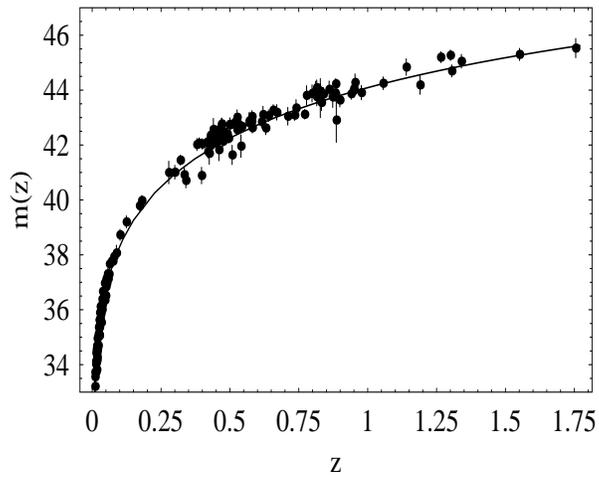}&%
\end{tabular}
 \caption{
 Magnitude $m$ as a function of red-shift $z$: Comparison of the theoretical curve to the observational data.}
\end{figure}

\begin{figure}[h!]
\centering
%\hspace{0.1cm}%
\begin{tabular}{cc}
\includegraphics*[width=240pt, height=180pt]{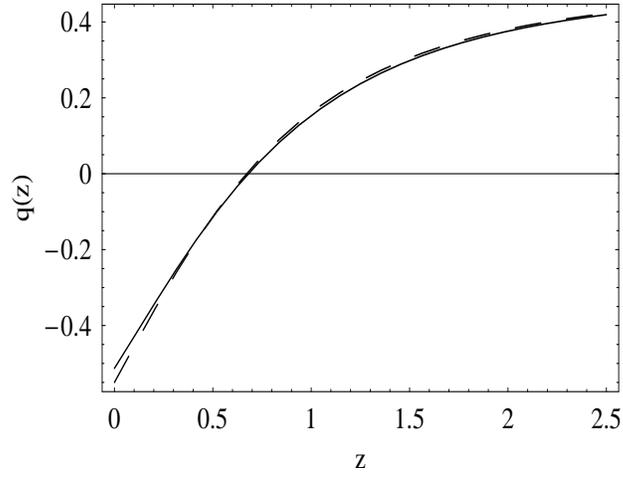}&%
\end{tabular}
 \caption{
 Deceleration parameter $q$ as a function of red-shift $z$ for LCDM model
 (dashed line) and our model (solid line).}
\end{figure}

\begin{figure}[h!]
\centering
%\hspace{0.1cm}%
\begin{tabular}{cc}
\includegraphics*[width=240pt, height=180pt]{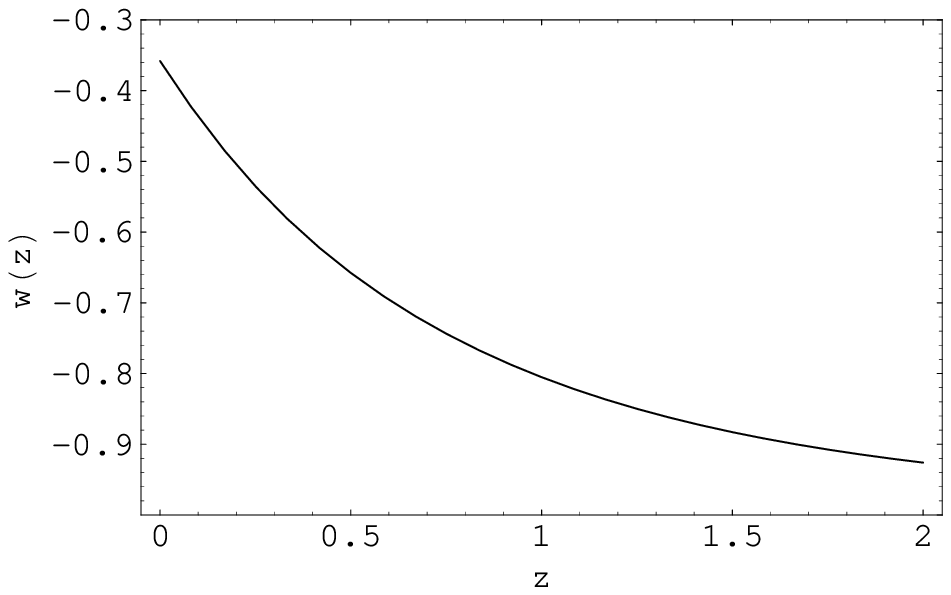}&%
\end{tabular}
 \caption{
Same as figure 1, but for the extreme case mentioned in the text.}
\end{figure}

\begin{figure}[h!]
\centering
%\hspace{0.1cm}%
\begin{tabular}{cc}
\includegraphics*[width=240pt, height=180pt]{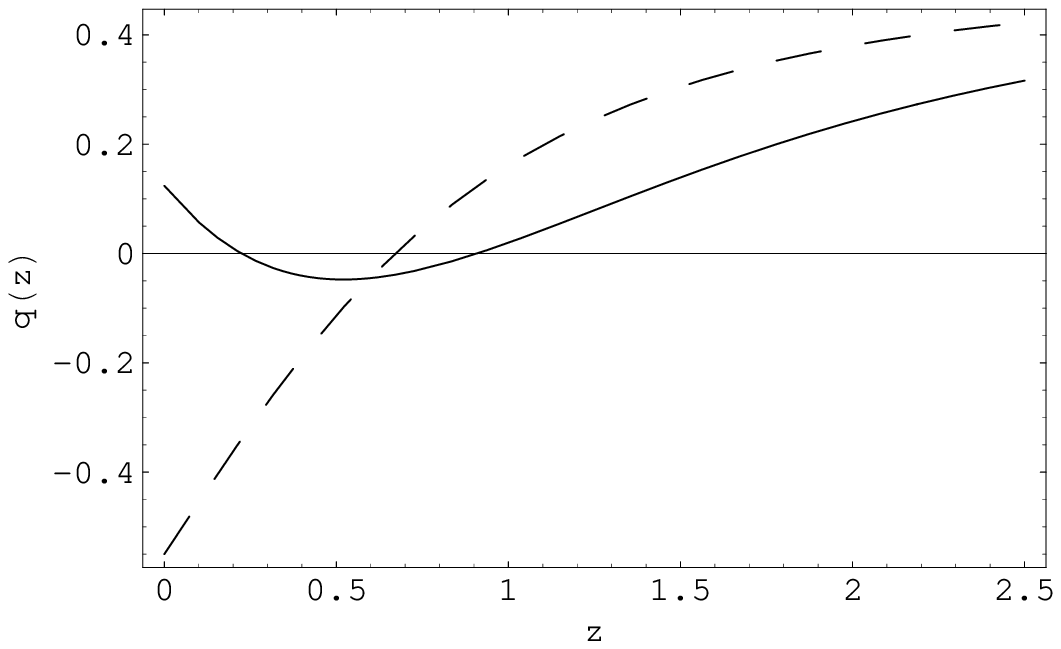}&%
\end{tabular}
 \caption{
Same as figure 3, but for the extreme case mentioned in the text.}
\end{figure}

\end{document}